\documentclass[aps,pra,showkeys,twocolumn,superscriptaddress]{revtex4-2}
\usepackage{amsmath, amssymb}
\usepackage{mathrsfs}
\usepackage{array}
\usepackage{booktabs}
\usepackage{tabu}
\usepackage{dcolumn}
\usepackage{amsmath}
\usepackage{amsfonts}
\usepackage{float}
\usepackage{amssymb}
\usepackage{graphicx,color}
\usepackage{tikz}
\usepackage[colorlinks={true}]{hyperref}
\hypersetup{colorlinks=true,linkcolor=red,citecolor=blue,urlcolor=blue}
\usepackage{graphicx}
\usepackage{subfigure}
\usepackage{graphicx}
\usepackage{dcolumn}
\usepackage{bm}
\usepackage{pstricks}
\usepackage{braket}
\usepackage{orcidlink}
\usepackage{comment}

\def\be{\begin{equation}}
\def\ee{\end{equation}}
\def\bea{\begin{eqnarray}}
\def\eea{\end{eqnarray}}
\def\f{\frac}
\def\n{\nonumber}
\def\l{\label}
\def\p{\phi}
\def\o{\over}
\def\R{\hat{\rho}}
\def\pa{\partial}
\def\om{\omega}
\def\na{\nabla}
\def\P{\Phi}
\begin{document} 
\title{Quantum Mpemba Effect in a Four-Site Bose--Hubbard Model
}
\author{Asad Ali\orcidlink{0000-0001-9243-417X}} \email{asal68826@hbku.edu.qa}
\affiliation{Qatar Center for Quantum Computing, College of Science and Engineering, Hamad Bin Khalifa University, Doha, Qatar}
\author{Hamid Arian Zad\orcidlink{0000-0002-1348-1777}}  
\email{hamid.arian.zad@upjs.sk}
\affiliation{Department of Theoretical Physics and Astrophysics, Faculty of Science of P. J. \v{S}af{\'a}rik University, Park Angelinum 9, 040 01 Ko\v{s}ice, Slovak Republic}
\author{Muhammad Irtiza Hussain\orcidlink{0000-0002-6231-7746}}
\affiliation{Qatar Center for Quantum Computing, College of Science and Engineering, Hamad Bin Khalifa University, Doha, Qatar}
\author{Saif Al-Kuwari\orcidlink{0000-0002-4402-7710}}
\email{smalkuwari@hbku.edu.qa}
\affiliation{Qatar Center for Quantum Computing, College of Science and Engineering, Hamad Bin Khalifa University, Doha, Qatar}
\author{Hashir Kuniyil\orcidlink{0000-0003-0338-1278}} 
\affiliation{Qatar Center for Quantum Computing, College of Science and Engineering, Hamad Bin Khalifa University,  Doha, Qatar}
\author{Muhammad Talha Rahim\orcidlink{0000-0003-1529-928X}} 
\affiliation{Qatar Center for Quantum Computing, College of Science and Engineering, Hamad Bin Khalifa University, Doha, Qatar}

\author{Michal Jaščur\orcidlink{0000-0003-0826-1961}}
\affiliation{Department of Theoretical Physics and Astrophysics, Faculty of Science of P. J. \v{S}af{\'a}rik University, Park Angelinum 9, 040 01 Ko\v{s}ice, Slovak Republic}
\author{Saeed Haddadi\orcidlink{0000-0002-1596-0763}} 
\email{haddadi@ipm.ir}
\affiliation{School of Particles and Accelerators, Institute for Research in Fundamental Sciences (IPM), P.O. Box 19395-5531, Tehran, Iran}

\date{\today}
\def\be{\begin{equation}}
\def\ee{\end{equation}}
\def\bea{\begin{eqnarray}}
\def\eea{\end{eqnarray}}
\def\f{\frac}
\def\n{\nonumber}
\def\l{\label}
\def\p{\phi}
\def\o{\over}
\def\R{\hat{\rho}}
\def\pa{\partial}
\def\om{\omega}
\def\na{\nabla}
\def\P{$\Phi$}

\begin{abstract}
We investigate relaxation-order inversion, known as the quantum Mpemba effect (QME), in a minimal open many-body system called a one-dimensional four-site Bose–Hubbard chain governed by Lindblad dynamics with local number dephasing. Families of thermal initial states are prepared at a fixed temperature and evolved under a common reference Liouvillian toward the same stationary state. Relaxation is characterized using four complementary diagnostics: trace distance, quantum relative entropy, symmetry-projected entropy imbalance (entanglement asymmetry), and the $\ell_{1}$-norm of coherence in the Fock basis. We find that QME emerges robustly in -the clean interacting regime, where on-site interactions redistribute the overlaps of initial states with slow Liouvillian decay modes, enabling states initially farther from equilibrium to converge faster at late times. In contrast, the noninteracting limit exhibits a monotonic relaxation hierarchy across all metrics. Introducing a linear Stark potential or random on-site disorder suppresses relaxation and eliminates QME signatures by inhibiting transport-assisted mixing and enhancing the dominance of slow modes. Within the explored parameter regime, the Stark field induces significantly stronger retardation than disorder. We further show that symmetry-projected entropy imbalance is particularly sensitive to charge-sector decoherence in reduced subsystems and provides a stringent probe of QME in bosonic platforms. Our results elucidate the essential role of interactions in enabling anomalous relaxation in open lattice systems and connect the suppression of QME under spatial inhomogeneity to localization phenomena in tilted and disordered Bose–Hubbard chains.
\end{abstract}


\maketitle

\section{Introduction}

The Mpemba effect refers to situations in which two initially distinct nonequilibrium states relaxing toward the same stationary state do so in a counterintuitive order. Namely, a state that is initially farther from equilibrium, according to a chosen distance measure, can approach equilibrium faster than another state that is initially closer. Importantly, temperature alone does not uniquely characterize the distance from equilibrium, and a colder system may in fact be farther from the stationary state than a hotter one. Originally discussed in the context of anomalous cooling processes~\cite{mpemba1969cool}, the Mpemba effect more generally reflects an inversion in the ordering of relaxation times in nonequilibrium dynamics~\cite{lu2017nonequilibrium,klich2019mpemba}. Its observation dates back to Aristotle~\cite{aristotle1952aristotle} and has since been reported in a wide variety of classical systems~\cite{ahn2016experimental,hu2018conformation,chaddah2010overtaking,greaney2011mpemba,lasanta2017hotter,keller2018quenches}, while continuing to stimulate debate and theoretical refinement~\cite{burridge2016questioning,bechhoefer2021fresh}.

A major theoretical advance emerged from stochastic thermodynamics, which clarified that relaxation dynamics are governed by the slowest decaying eigenmodes of the dynamical generator~\cite{lu2017nonequilibrium,klich2019mpemba}. Within this framework, the Mpemba effect arises when initial states that are farther from equilibrium exhibit smaller overlap with these slow modes than states initially closer to equilibrium~\cite{PhysRevLett.134.107101,summer2025}. In extreme cases, termed the strong Mpemba effect, this overlap vanishes entirely, leading to an exponential acceleration of relaxation~\cite{klich2019mpemba}.
In recent years, there has been growing interest in quantum analogues of this phenomenon, known as the quantum Mpemba effect (QME). In quantum systems, relaxation can depend sensitively on coherence, correlations, symmetry structure, and environmental coupling. Two broad classes of QME have been identified. The first concerns open quantum systems described by Lindblad master equations, where relaxation is controlled by the Liouvillian spectrum and by the overlaps of initial states with slow decay modes~\cite{carollo2021exponentially,longhi2025quantum,PhysRevB.111.104312}. The second involves closed quantum systems following quenches, in which relaxation and symmetry restoration are governed by unitary dynamics and can be probed through symmetry-resolved entropies and related observables~\cite{rylands2024dynamical,liu2024symmetry}. Together, these settings provide a unifying framework connecting classical-like dissipation with intrinsically quantum relaxation mechanisms.

Several complementary mechanisms have been proposed to account for QME. Besides the strong Mpemba effect associated with suppressed overlap with slow Liouvillian modes~\cite{carollo2021exponentially,zhang2025observation}, non-normal dynamics can generate transient interference between decay channels~\cite{longhi2025quantum}. Initial  correlations between system and environment, and non-Markovian memory effects may further accelerate equilibration~\cite{ares2025quantum}. In closed systems, symmetry restoration and entanglement asymmetry dynamics can drive anomalous relaxation~\cite{yu2025tuning,ares2025quantum}, while integrability and localization phenomena introduce additional relaxation pathways distinct from chaotic systems~\cite{ares2025quantum,longhi2025mpemba,PhysRevLett.133.010401}.

These theoretical developments have been complemented by recent experimental breakthroughs. Joshi \emph{et al.} observed QME in a trapped-ion quantum simulator using entanglement asymmetry as a probe of symmetry restoration~\cite{joshi2024observing}. Zhang \emph{et al.} demonstrated the first unambiguous realization of the strong QME in a single-ion platform~\cite{zhang2025observation}. An inverse QME, in which a colder state equilibrates faster than a warmer one, has also been reported experimentally~\cite{aharony2024inverse}. These advances highlight the need for systematic investigations of QME in interacting and spatially structured quantum systems.

Following the discussion of localization phenomena in tilted and disordered Bose--Hubbard chains, early studies by A. Tomadin {\it et al.} \cite{Tomadin2007} and P. Buonsante {\it et al.} \cite{Buonsante2008} established that spatial inhomogeneities fundamentally alter many-body dynamics, where the former demonstrating interaction-induced destruction of Bloch oscillations in tilted lattices, and the latter revealing a disorder-driven crossover from regular to quantum-chaotic dynamics. These works highlight the competition between interactions, which promote delocalization and mixing, and spatial inhomogeneities, which suppress transport and enhance localization. Our study extends this understanding to the open-system context, showing that the same localization mechanisms that inhibit coherent transport also suppress the anomalous relaxation ordering characteristic of the QME.

Motivated by this progress, we investigate QME in a  four-site Bose--Hubbard model on a linear chain subject to Markovian dephasing. The Bose--Hubbard model exhibits a superfluid--Mott insulator transition driven by the competition between tunneling and interactions \cite{fisher1989boson,jaksch1998cold,bloch2008many}.
 We address two central questions: (i) how on-site interactions enable or suppress relaxation-order inversion, and (ii) how spatial inhomogeneities, specifically a linear Stark field and random on-site disorder, modify the slow relaxation modes and influence QME signatures. 
 While our numerical analysis focuses on a four-site, unit-filled chain that allows exact treatment of the full Liouvillian dynamics, we emphasize physical mechanisms based on mode structure, transport constraints, and localization effects that are expected to generalize qualitatively.
In addition, we examine how the manifestation of QME depends on the choice of distance measure from equilibrium. We employ multiple metrics, including trace distance, quantum relative entropy, entanglement asymmetry, and coherence-based measures, to characterize the robustness and universality of anomalous relaxation behavior.

This paper is organized as follows. Section~\ref{sec:model} introduces the model, the Lindblad description, and the metrics used to detect QME. Section~\ref{sec:method} details the numerical protocol defining a common dynamical generator and families of initial conditions. Section~\ref{sec:result} presents the results in clean, interacting, Stark, and disordered regimes, with interpretation in terms of Liouvillian mode hierarchy and localization-induced relaxation suppression. Section~\ref{sec:conclusion} summarizes our findings and outlines directions for future work.

\section{Model and Methods}\label{sec:model}
\subsection{The model}
The Bose--Hubbard Hamiltonian describes the relaxation dynamics of interacting bosons on a one-dimensional lattice:
\begin{equation}
\begin{aligned}
\hat{H}_{\mathrm{BH}} ={} & -\tau \sum_{j=1}^{M-1} \left( \hat{a}_j^\dagger \hat{a}_{j+1} + \hat{a}_{j+1}^\dagger \hat{a}_j \right) \\
& - \mu \sum_{j=1}^{M} \hat{n}_j + \frac{U}{2} \sum_{j=1}^{M} \hat{n}_j (\hat{n}_j - 1),
\end{aligned}
\label{eq:BHM}
\end{equation}
where $M$ is the number of lattice sites, $\hat{a}_j^\dagger$ ($\hat{a}_j$) creates (annihilates) a boson at site $j$, and $\hat{n}_j = \hat{a}_j^\dagger \hat{a}_j$ is the number operator. The total number of particles is denoted by $N$. In this study, we focus on the unit-filling regime characterized by $N = 4$ bosons on a four-site lattice with $M = 4$.
Parameters $\tau$, $\mu$, and $U$ denote the tunneling amplitude, chemical potential, and on-site interaction strength, respectively. The competition between tunneling and interactions drives a quantum phase transition between superfluid and Mott insulator phases. To incorporate environmental effects and external fields, we extend the Hamiltonian as
\begin{equation}
\hat{H} = \hat{H}_{\mathrm{BH}} + \sum_{j=1}^{M} \left( g j + \delta_j \right) \hat{n}_j,
\label{eq:full_H}
\end{equation}
where $g$ is the strength of the linear Stark field and $\delta_j$ represents a site-dependent disorder uniformly distributed in $[-\delta, +\delta]$ \cite{ali2025coherence}.

\subsection{Liouvillian mode picture and relation to QME}
\label{subsec:liouvillian_picture}

A microscopic derivation of the Lindblad master equation follows from weak system--bath coupling, a separation of bath and system timescales, and the Born--Markov and secular approximations, yielding a time-homogeneous completely positive trace-preserving (CPTP) semigroup generator \cite{breuer2002theory, gorini1976completely, lindblad1976generators}. In cold-atom implementations, local dephasing can arise from off-resonant light scattering, phase noise in optical potentials, or coupling to uncontrolled background fields, and is commonly modeled by jump operators proportional to local densities \cite{gopalakrishnan2011universal,barontini2013controlling}. Interaction between the system and an external environment causes decoherence which under the Markovian approximation cab be modeled through the Lindblad master equation:
\begin{align}
	\frac{d\hat{\rho}(t)}{dt} 
	&= \mathcal{L}[\hat{\rho}(t)] \nonumber \\
	&= -i[\hat{H}, \hat{\rho}(t)] 
	+ \sum_k \left( 
	\hat{L}_k \hat{\rho}(t) \hat{L}_k^\dagger 
	- \frac{1}{2} \{ \hat{L}_k^\dagger \hat{L}_k, \hat{\rho}(t) \} 
	\right),
	\label{eq:lindblad}
\end{align}
where $\mathcal{L}$ is the Liouvillian superoperator and $\hat{L}_k$ are Lindblad operators characterizing the system-environment interaction. The steady state $\hat{\rho}_{\mathrm{ss}}$ satisfies $\mathcal{L}[\hat{\rho}_{\mathrm{ss}}] = 0$. We consider dephasing as the primary dissipative mechanism, modeled by $\hat{L}_j = \sqrt{\gamma} \, \hat{n}_j$, where $\gamma$ is the dephasing rate acting on the local particle number at each site $j$, reflecting realistic experimental conditions.

The dynamics of an open quantum system governed by a time-independent Lindbladian generator $\mathcal{L}$ admits a spectral decomposition in terms of its eigenmodes. The density matrix evolution can be expressed as
\begin{equation}
	\hat{\rho}(t) = \hat{\rho}_{\mathrm{ss}} + \sum_{j=1}^{N-1} c_j e^{\lambda_j t} \hat{\rho}_j,
	\label{eq:spectral_decomp}
\end{equation}
where $\hat{\rho}_{\mathrm{ss}}$ denotes the unique steady state, $\lambda_j$ are the nonzero eigenvalues of $\mathcal{L}$ with $\mathrm{Re}(\lambda_j)<0$, $\hat{\rho}_j$ are the corresponding right eigenmatrices, and the coefficients $c_j$ depend on the initial state $\hat{\rho}(0)$. The long-time relaxation dynamics is governed by the eigenvalues closest to the imaginary axis, commonly referred to as the slow modes. The characteristic relaxation time is set by the Liouvillian gap
$\Delta_{\mathcal{L}}=\min_{j\ge1}|\mathrm{Re}(\lambda_j)|$,
while the presence of additional low-lying modes can generate metastable plateaus
\cite{macieszczak2016towards,zhang2025observationJ,brown2023unravelling}.
The eigenmode with the largest real part ultimately determines the asymptotic approach to the steady state.

Within this Liouvillian mode picture, the QME emerges naturally as a consequence of the redistribution of initial-state overlaps onto the slow subspace. The QME describes a non-equilibrium phenomenon in which a system initialized farther from equilibrium relaxes faster than one prepared closer to equilibrium. Formally, consider two initial states $\hat{\rho}_1(0)$ and $\hat{\rho}_2(0)$ evolving toward $\hat{\rho}_{\mathrm{ss}}$. The QME occurs when there exists a crossing time $t_M>0$ such that for all $t>t_M$ we have 
$d(\hat{\rho}_1(0),\hat{\rho}_{\mathrm{ss}})> (\hat{\rho}_2(0),\hat{\rho}_{\mathrm{ss}}),
\quad$ {but} $ d(\hat{\rho}_1(t),\hat{\rho}_{\mathrm{ss}})
<d(\hat{\rho}_2(t),\hat{\rho}_{\mathrm{ss}})$,
where $d(\cdot,\cdot)$ denotes a suitable distance measure on the space of quantum states. In terms of the spectral decomposition, such a reordering arises when the initially more distant state has a reduced overlap with the slowest-decaying eigenmodes. As a result, it relaxes more rapidly at late times, leading to an inversion of relaxation ordering.

A particularly striking manifestation is the strong Mpemba effect, which occurs when one initial state has a vanishing or strongly suppressed projection onto the slowest mode. In this case, relaxation proceeds exponentially faster than for generic initial conditions. Such behavior can be systematically engineered through appropriate state preparation protocols
\cite{carollo2021exponentially,zhang2025observation}.
In interacting many-body systems, on-site interactions can significantly reshape both the slow-mode structure and the mapping between thermal initial conditions and the coefficients $c_j$, thereby enabling relaxation-order inversion. By contrast, strong Stark gradients or disorder suppress transport and mixing, enhancing slow-mode dominance and inhibiting metric crossings within accessible parameter regimes.

The quantitative characterization of QME depends sensitively on the choice of distance measure. A widely used geometric metric is the trace distance,
\begin{equation}
	D[\hat{\rho}(t),\hat{\rho}_{\mathrm{ss}}]
	=
	\frac{1}{2}
	\mathrm{Tr}\!\left[
	\sqrt{
		(\hat{\rho}(t)-\hat{\rho}_{\mathrm{ss}})^\dagger
		(\hat{\rho}(t)-\hat{\rho}_{\mathrm{ss}})
	}
	\right],
\end{equation}
which is contractive under completely positive trace-preserving maps, bounded between $0$ and $1$, and admits a clear operational interpretation in quantum hypothesis testing
\cite{nielsen2010quantum}.

From a thermodynamic viewpoint, the quantum relative entropy provides an information-theoretic measure of nonequilibrium,
\begin{equation}
	S[\hat{\rho}(t)\Vert\hat{\rho}_{\mathrm{ss}}]
	=
	\mathrm{Tr}\!\left[
	\hat{\rho}(t)
	\big(
	\log\hat{\rho}(t)-\log\hat{\rho}_{\mathrm{ss}}
	\big)
	\right],
\end{equation}
which is directly related to nonequilibrium free energy differences and satisfies monotonicity under quantum channels
\cite{moroder2024thermodynamics}.

Beyond global metrics, local and symmetry-resolved probes provide complementary insights into thermalization dynamics. Entanglement asymmetry serves as a sensitive diagnostic of symmetry restoration. For a subsystem $A$ and its complement $\overline{A}$, we define the R\'enyi entanglement asymmetry of order $n$ as
\begin{equation}
	\Delta S_A^{(n)}(t)
	=
	S^{(n)}(\rho_{A,Q}(t))
	-
	S^{(n)}(\rho_A(t)),
\end{equation}
where $\rho_A(t)=\mathrm{Tr}_{\overline{A}}[\hat{\rho}(t)]$ and
\begin{equation}
	\rho_{A,Q}(t)
	=
	\sum_q \Pi_q \rho_A(t) \Pi_q
\end{equation}
is the charge-symmetrized reduced density matrix. Here, $\Pi_q$ projects onto the eigenspace of the conserved charge
$Q_A=\sum_{i\in A}\hat{n}_i$
with eigenvalue $q$, and
$S^{(n)}(\rho)=(1-n)^{-1}\log\mathrm{Tr}(\rho^n)$
denotes the R\'enyi entropy. In this work, we focus on the von Neumann limit $n=1$. The entanglement asymmetry vanishes when $\rho_A(t)$ is block-diagonal in the charge sectors, thereby signaling symmetry restoration
\cite{joshi2024observing}.

Another important indicator of relaxation is the $\ell_1$-norm of quantum coherence,
\begin{equation}
	\mathcal{C}[\hat{\rho}(t)]
	=
	\sum_{i\neq j}
	\left|
	\langle i|\hat{\rho}(t)|j\rangle
	\right|,
\end{equation}
where $\{|i\rangle\}$ denotes the Fock basis. This quantity captures the total amount of superposition in a preferred basis and has been shown to be sensitive to dynamical phase transitions and thermalization scaling
\cite{Baumgratz2014}.

Taken together, these metrics reveal complementary facets of the QME. While trace distance and relative entropy quantify the global approach to equilibrium, coherence measures probe superposition decay, and entanglement asymmetry monitors symmetry restoration in subsystems. The strong Mpemba effect is often most pronounced in entanglement-related observables, where initially more symmetry-broken states may exhibit faster relaxation
\cite{rylands2024dynamical,liu2024symmetry}.
Experimentally, these quantities can be accessed using quantum state tomography
\cite{huang2020predicting}
or randomized measurement techniques
\cite{brydges2019probing},
enabling direct observation of QME in trapped-ion and related platforms
\cite{joshi2024observing}.

The dependence of QME on the chosen metric highlights the multifaceted nature of quantum thermalization. Different measures may exhibit crossings at distinct times or even qualitatively different behavior, emphasizing the importance of selecting observables tailored to specific experimental and theoretical settings
\cite{ares2025quantum,moroder2024thermodynamics}.
Consequently, robust identification of QME typically relies on the combined analysis of multiple complementary measures.

In this study, we focus on the particular chemical potential at $\mu = 0.5$, corresponding to the center of the first Mott lobe in the Bose--Hubbard phase diagram. This choice yields a Hilbert space of dimension $D=35$, while the associated Liouvillian superoperator has dimension $D^2=1225$. Owing to the combinatorial growth of the state space and the computational cost of solving the full Lindblad master equation, we restrict our analysis to this minimal yet nontrivial setting. Despite its modest size, this system retains the essential ingredients required to capture symmetry-breaking dynamics and dissipative effects in a controlled and numerically exact manner. We declare that a full spectral analysis is, in principle, feasible. Due to space constraints, we will pursue this exploration in a forthcoming work to extract $\Delta_{\mathcal{L}}$ and the low-lying modes as functions of the parameters.

\section{Procedure for Probing QME}
\label{sec:method}

A necessary condition for a meaningful Mpemba comparison is that all candidate initial states evolve under the \emph{same} dynamical generator. Throughout this work, we therefore ensure that all relaxation processes are governed by a fixed reference Liouvillian $\mathcal{L}_{\rm ref}$ with a unique stationary state $\rho_{ss}$ satisfying
$\mathcal{L}_{\rm ref}[\rho_{ss}]=0$.
This guarantees that any observed relaxation-order inversion arises solely from differences in the initial conditions rather than from variations in the dynamics.

To systematically investigate the emergence and characteristics of the QME in the four-site Bose--Hubbard model, we employ a controlled numerical protocol that isolates the roles of kinetic energy, interactions, and spatial inhomogeneity under dissipative Lindblad evolution. The core strategy consists of preparing a hierarchy of thermal initial states with varying distances from a common steady state and tracking their relaxation using multiple complementary quantum metrics.

\subsection{System Specification and Parameter Regimes}

We consider a four-site lattice at unit filling with $N=4$ bosons and chemical potential $\mu=0.5$. The system Hamiltonian is given in Eq.~\eqref{eq:full_H}. We explore four representative physical regimes by selectively tuning model parameters:

\begin{itemize}
	\item \textbf{Clean, Non-interacting:} $U=0$, $g=0$, $\delta=0$, with hopping $\tau$ as the control parameter.
	\item \textbf{Clean, Interacting:} $U>0$ (fixed), $g=0$, $\delta=0$, with $\tau$ as the control parameter.
	\item \textbf{Stark Potential:} $U>0$ (fixed), $\tau$ (fixed, low), $\delta=0$, with Stark field strength $g$ as the control parameter.
	\item \textbf{Random Disorder:} $U>0$ (fixed), $\tau$ (fixed, low), $g=0$, with disorder amplitude $\delta$ as the control parameter.
\end{itemize}

For each regime, a reference value $\theta_{\rm ref}$ of the control parameter is chosen to define the common generator $\mathcal{L}_{\rm ref}$ and the associated steady state $\rho_{ss}$.

\subsection{Initial State Preparation and Common Generator}

For a given regime, we prepare a family of thermal initial states by varying a preparation parameter $\theta$ in the Hamiltonian used solely for state preparation. Each initial state is given by the Gibbs ensemble
\begin{equation}
	\rho_\beta^{(\theta)} =
	\frac{e^{-\beta H(\theta)}}{\mathrm{Tr}(e^{-\beta H(\theta)})},
\end{equation}
where $\beta$ is the inverse temperature and $H(\theta)$ denotes the Hamiltonian with control parameter $\theta\in\{\tau,g,\delta\}$.

While the preparation Hamiltonian is varied, all states are evolved under the \emph{same} reference Liouvillian:
\begin{equation}
	\dot{\rho}(t)=\mathcal{L}_{\rm ref}[\rho(t)], \qquad
	\rho(0)=\rho_\beta^{(\theta)}.
\end{equation}
This procedure generates a controlled hierarchy of initial conditions with systematically varying physical properties, while keeping the target state and the dynamical map identical across all runs.

\subsection{Time Evolution and Quantification of Relaxation}

Each initial state is evolved according to the Lindblad master equation~\eqref{eq:lindblad} with local dephasing operators
$\hat{L}_j=\sqrt{\gamma}\hat{n}_j$
and fixed dephasing rate $\gamma$. The equation is solved numerically to obtain the exact time-dependent density matrix $\rho(t)$
\cite{lambert2024,johansson2013qutip,johansson2012qutip}.

The distance from the steady state is monitored using several complementary metrics:

\begin{itemize}
	\item \textbf{Trace distance.}
	$D[\rho(t),\rho_{ss}]$, quantifying geometric distinguishability.
	
	\item \textbf{Relative entropy.}
	$S[\rho(t)\Vert\rho_{ ss}]$, measuring information-theoretic divergence.
	
	\item \textbf{Entanglement asymmetry} (symmetry-projected entropy imbalance).
	We employ a symmetry-sensitive diagnostic inspired by symmetry-resolved entanglement studies \cite{joshi2024observing}. For a bipartition into subsystem $A$ and complement $\overline{A}$, define the reduced state $\rho_A(t)=\text{Tr}_{\overline{A}}[\rho(t)]$. Let $Q_A=\sum_{i\in A}\hat n_i$ be the particle-number operator on $A$ with projectors $\{\Pi_q\}$ onto its eigenspaces. We define the dephased (charge-projected) reduced state
	$\rho_{A,Q}(t)=\sum_q \Pi_q\,\rho_A(t)\,\Pi_q$,
	and consider the entropy difference
	$\Delta S_A(t)=S(\rho_{A,Q}(t))-S(\rho_A(t))$,
	where $S(\rho)=-\text{Tr}(\rho\log\rho)$.
	For globally pure states, $\Delta S_A$ coincides with a symmetry-resolved entanglement asymmetry. In our open-system setting, $\rho(t)$ is mixed (purity decreases monotonically under Hermitian dephasing), hence $\Delta S_A$ should not be interpreted as an entanglement monotone. Rather, it quantifies the extent to which $\rho_A(t)$ contains coherence between different $Q_A$ sectors: $\Delta S_A=0$ if and only if $\rho_A(t)$ is block-diagonal in the $Q_A$ basis. It therefore acts as a sharp indicator of charge-sector decoherence and symmetry-sector restoration during relaxation, and has proven experimentally accessible in related trapped-ion settings \cite{joshi2024observing}.
	
	\item \textbf{$\ell_1$-norm of quantum coherence.}
$	\mathcal{C}[\rho(t)] = \sum_{i\neq j}
		\left|\langle i|\rho(t)|j\rangle\right|$,
	tracking the decay of quantum superpositions.
\end{itemize}

Together, these observables provide a multi-faceted characterization of the relaxation dynamics.

\subsection{Identification of the QME}

The QME is identified through characteristic crossings in the time evolution of the chosen distance measures. For two initial states $\rho_\beta^{(\theta_1)}$ and $\rho_\beta^{(\theta_2)}$, the effect is established if
\begin{equation}
	d(\rho_\beta^{(\theta_1)},\rho_{ss})
	>
	d(\rho_\beta^{(\theta_2)},\rho_{ss})
	\quad \text{at } t=0,
\end{equation}
but there exists a finite time $t_M>0$ such that for $t>t_M$,
\begin{equation}
	d(\rho^{(\theta_1)}(t),\rho_{ss})
	<
	d(\rho^{(\theta_2)}(t),\rho_{ss}).
\end{equation}
Here, $d(\cdot,\cdot)$ denotes any of the employed metrics.

Robust observation of the QME is confirmed by consistent crossover behavior across multiple measures. This unified protocol enables a controlled and comparative analysis of how different physical parameters influence anomalous relaxation, thereby identifying the essential ingredients underlying the QME.

\section{Results and Discussion}\label{sec:result}
\subsection{Clean Tight-Binding Regime}
\label{subsec:clean_tb}

Following the general procedure outlined in Sec.~\ref{sec:method}, we initiate our investigation of the QME by examining the clean non-interacting Bose--Hubbard model, setting on-site interaction, disorder, and Stark potential to zero ($U = \delta = g = 0$) with a chemical potential ensuring unit filling. This configuration isolates purely kinetic effects, serving as a baseline to elucidate the role of interactions in quantum thermalization dynamics.

We implement the procedure using a four-site bosonic chain with four particles. Initial thermal states are prepared at a fixed temperature, each characterized by a distinct hopping strength, and evolved under Lindblad dynamics with local dephasing noise acting on particle number operators. All systems converge toward a common reference steady state defined by a specific hopping strength. This approach establishes a hierarchy of initial states with varying degrees of delocalization while preserving identical thermal properties, isolating quantum effects from classical thermal phenomena. Relaxation dynamics are tracked using four complementary quantum metrics: trace distance $D[\hat{\rho}_{ss}, \hat{\rho}(t)]$ to quantify geometric distinguishability, relative entropy $S[\hat{\rho}(t) || \hat{\rho}_{ss}]$ for information-theoretic divergence, entanglement asymmetry $\Delta S_A$ for a chosen bipartition to probe particle number symmetry restoration, and $\ell_1$-norm of quantum coherence, $\mathcal{C}[\hat{\rho}(t)]$, in the Fock basis to capture quantum superposition dynamics.

\begin{figure*}[t]
\centering
\includegraphics[width=0.80\linewidth]{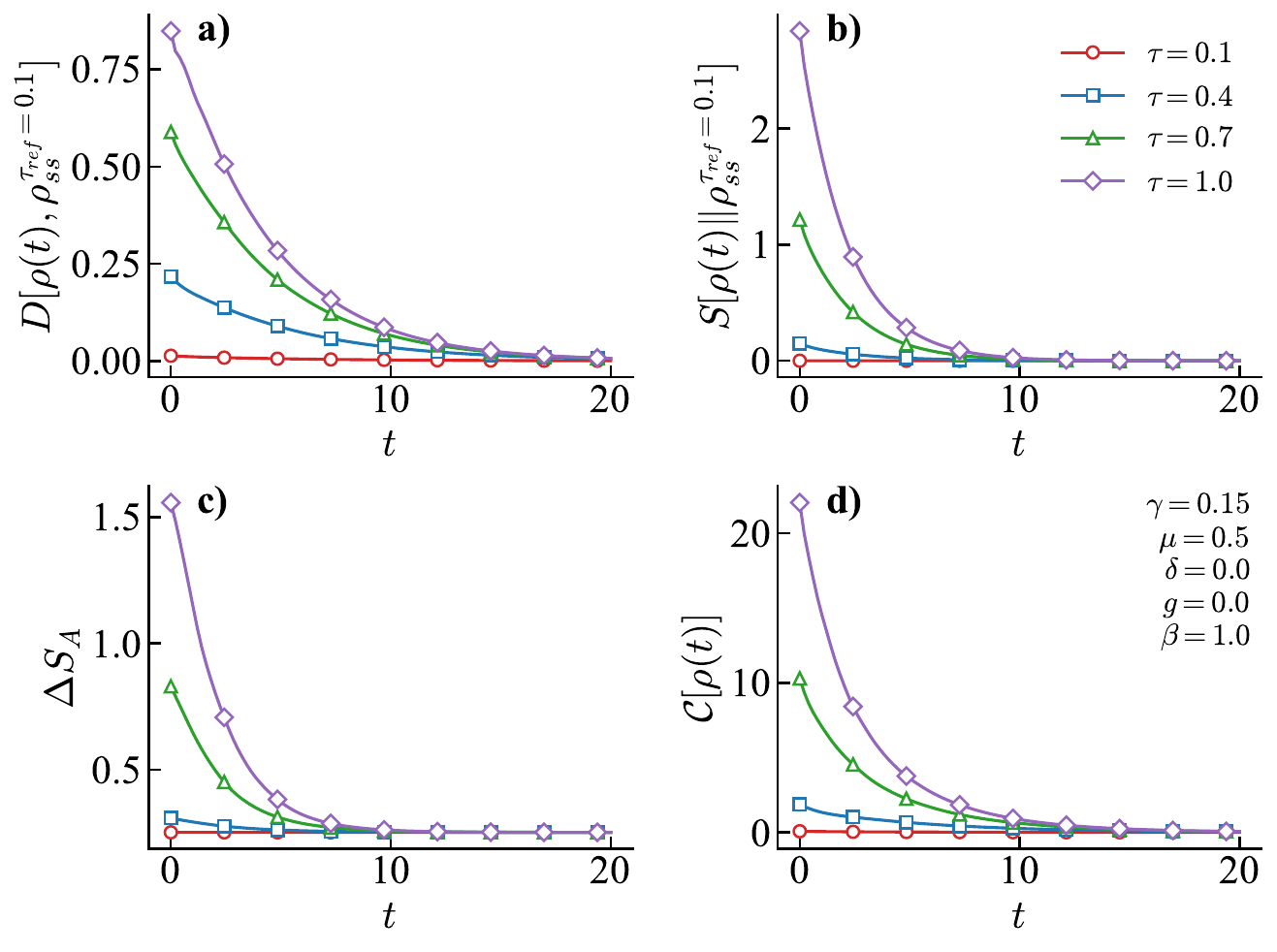}
\caption{Absence of the QME in the non-interacting regime ($U=0$). Relaxation dynamics of a dissipative bosonic system under varying hopping strengths $\tau$. A four-site Bose--Hubbard model evolves from thermal initial states under Lindblad dynamics with local dephasing noise (rate $\gamma$). The reference steady state $\hat{\rho}_{ss}$ corresponds to a fixed hopping strength $\tau_\text{ref}$. All four metrics show consistent monotonic relaxation without crossovers: \textbf{(a)} trace distance $D[\hat{\rho}(t), \hat{\rho}_{ss}]$ measuring distinguishability, \textbf{(b)} relative entropy $S[\hat{\rho}(t) || \hat{\rho}_{ss}]$ quantifying information-theoretic divergence, \textbf{(c)} entanglement asymmetry $\Delta S_A$ for bipartition $A$ characterizing particle number symmetry breaking, and \textbf{(d)} $\ell_1$-norm of quantum coherence $\mathcal{C}[\hat{\rho}(t)]$ in the Fock basis. States initially closer to the steady state (lower $\tau$) maintain their proximity advantage throughout evolution, demonstrating conventional thermalization and highlighting the necessity of interactions for QME emergence. 
}
\label{fig:relaxation_dynamics}
\end{figure*}

In the non-interacting limit the preparation protocol changes primarily single-particle delocalization but does not generate interaction-induced redistribution among slow decay modes. Under local number dephasing, coherences in the Fock basis decay rapidly on timescale $\sim\gamma^{-1}$, while populations mix via the Hamiltonian. In this regime the ordering of overlaps with the slow subspace remains aligned with the initial metric ordering, yielding conventional monotone relaxation.
The results shown in Fig.~\ref{fig:relaxation_dynamics} reveal the absence of QME signatures in the non-interacting regime. All quantum metrics exhibit monotonic ordering throughout the evolution, with initial states closer to the reference steady state consistently maintaining their proximity advantage. Specifically, the trace distance in panel~\ref{fig:relaxation_dynamics}(a) displays a clear hierarchy, with no crossover events indicative of anomalous relaxation. This ordering persists across relative entropy [panel~\ref{fig:relaxation_dynamics}(b)], entanglement asymmetry [panel~\ref{fig:relaxation_dynamics}(c)], and quantum coherence [panel~\ref{fig:relaxation_dynamics}(d)], confirming conventional thermalization where initial proximity to equilibrium dictates relaxation speed.

\vspace{-0.07cm}
Notably, entanglement asymmetry indicates that systems with higher hopping strengths exhibit greater initial symmetry breaking, which persists throughout relaxation, suggesting that symmetry-breaking content hinders convergence to the symmetric steady state. The absence of crossover behavior establishes that kinetic effects alone are insufficient to induce QME, indicating the necessity of interactions for anomalous relaxation dynamics in this model.

\subsection{Clean Bose--Hubbard Regime}
\label{subsec:clean_bhm}

Building on the non-interacting case discussed in Sec.~\ref{subsec:clean_tb}, we extend our analysis by introducing on-site interactions in the clean Bose--Hubbard model. Following the procedure outlined in Sec.~\ref{sec:method}, we set the disorder and Stark potentials to zero ($\delta = g = 0$) and fix the chemical potential to ensure unit filling. This allows us to isolate the effects of many-body correlations on quantum thermalization and to contrast them with the purely kinetic dynamics of the non-interacting regime.
The on-site interaction term penalizes multiple occupancy, reshaping the energy landscape and generating competition between kinetic delocalization induced by hopping and interaction-driven localization. At unit filling, this competition enhances sensitivity to correlation effects, making this regime particularly suitable for probing interaction-driven quantum Mpemba behavior.
Thermal initial states are prepared at a fixed inverse temperature for different hopping strengths and evolved under Lindblad dynamics with local dephasing noise acting on particle number operators, as in Sec.~\ref{subsec:clean_tb}. All states relax toward a common reference steady state defined by a fixed hopping strength $\tau_{\rm ref}$. The relaxation dynamics are characterized using the four described quantum metrics in the Fock basis.
\begin{figure*}[t]
	\centering
	\includegraphics[width=0.80\linewidth]{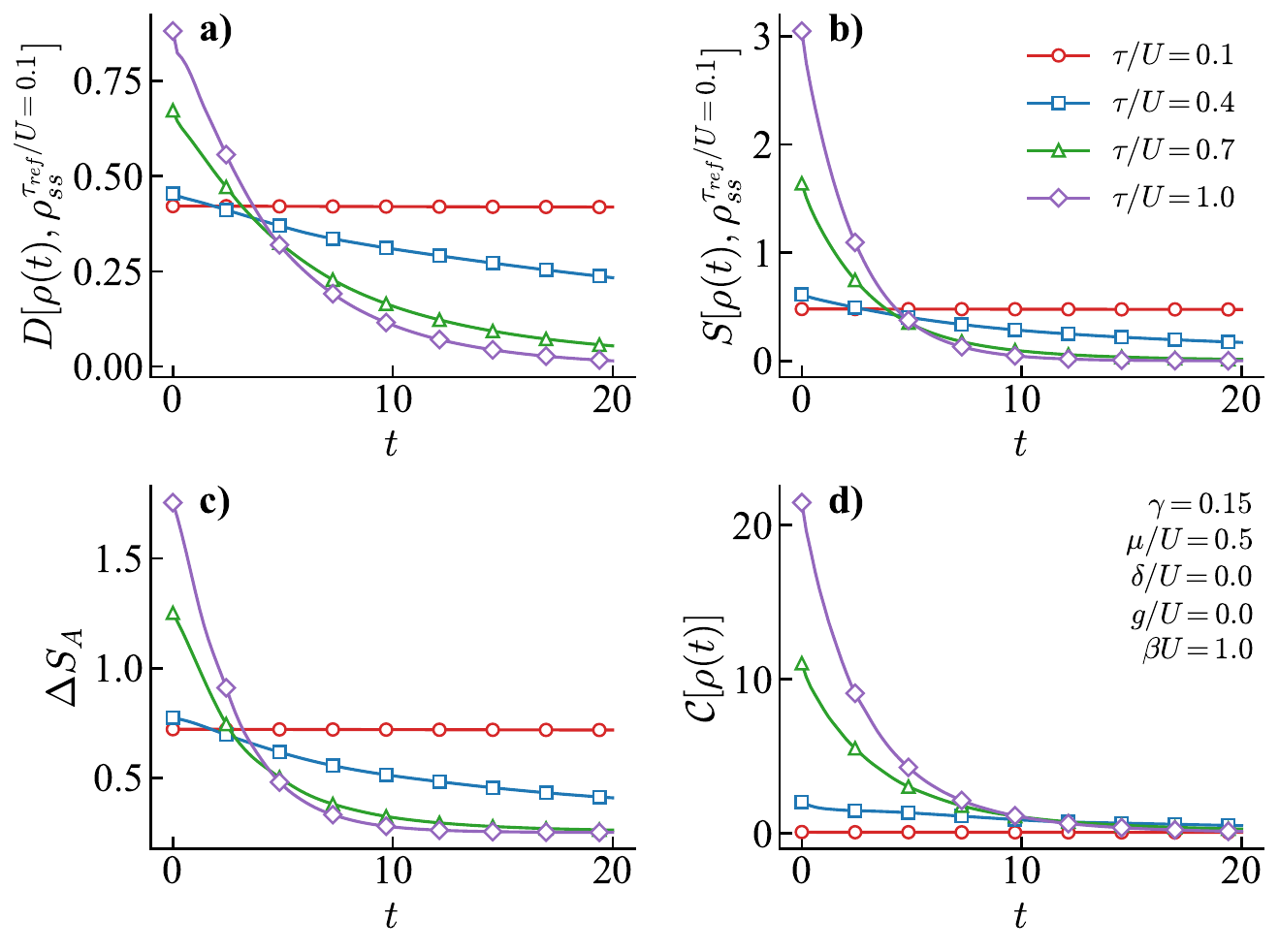}
	\caption{Emergence of the QME in the interacting regime. Relaxation dynamics of a dissipative bosonic chain with finite on-site interactions under varying hopping strengths $\tau$. A four-site system evolves from thermal initial states under Lindblad dynamics with local dephasing noise at rate $\gamma$. The reference steady state $\hat{\rho}_{ss}$ corresponds to a fixed hopping strength $\tau_{\rm ref}$. All four diagnostics exhibit characteristic crossover behavior indicative of the QME: (a) trace distance $D[\hat{\rho}(t), \hat{\rho}_{ss}]$, showing inversion of the relaxation hierarchy; (b) relative entropy $S[\hat{\rho}(t)|\hat{\rho}_{ss}]$, demonstrating faster convergence for initially more distant states; (c) entanglement asymmetry $\Delta S_A$ for bipartition $A$, revealing enhanced symmetry-restoration dynamics; and (d) $\ell_1$-norm of quantum coherence $\mathcal{C}[\hat{\rho}(t)]$ in the Fock basis, highlighting interaction-modified decoherence pathways. States initially farther from equilibrium (larger $\tau$) overtake closer ones, demonstrating anomalous thermalization enabled by many-body interactions. 
    }
\label{fig:interacting_dynamics}
\end{figure*}
Figure~\ref{fig:interacting_dynamics} demonstrates a striking departure from the non-interacting case, where no QME was observed. The introduction of on-site interactions induces clear quantum Mpemba signatures across all four metrics. In panel~\ref{fig:interacting_dynamics}(a), the trace distance exhibits multiple crossover events, where the systems initially farther from equilibrium overtake those that start closer, indicating an inversion of the relaxation hierarchy. Panel~\ref{fig:interacting_dynamics}(b) shows a similar ordering reversal in the relative entropy, where states with larger initial information-theoretic distance converge more rapidly at later times. This behavior contrasts sharply with the monotonic relaxation observed in Sec.~\ref{subsec:clean_tb} and confirms the emergence of anomalous thermalization.
 Remaining two quantities provide further insight into the microscopic mechanism. The entanglement asymmetry shown in panel~\ref{fig:interacting_dynamics}(c) reveals that interactions enable efficient restoration of particle-number symmetry, with initially more asymmetric states reaching lower final asymmetry. Meanwhile, the evolution of quantum coherence (panel~\ref{fig:interacting_dynamics}(d)) demonstrates that on-site interactions restructure decoherence pathways, allowing access to superposition states that connect more efficiently to the steady state. Together, these effects facilitate the counterintuitive relaxation behavior characteristic of the QME.

In the interacting regime, all four diagnostics exhibit clear crossings. This indicates that the interaction-modified many-body structure reshapes the coefficient vector ${c_\alpha}$ in the Liouvillian expansion. In particular, certain thermal initial states prepared at larger $\tau/U$ carry reduced weight on the slowest-decaying modes of $\mathcal{L}_{\rm ref}$, despite being initially farther from $\hat{\rho}_{ss}$ in standard distance measures. As a result, these states converge faster at late times and overtake initially closer states, producing the observed inversion. The simultaneous appearance of crossings in distinguishability, thermodynamic divergence, charge-sector coherence, and basis coherence confirms that the effect is not an artifact of any single metric.

Overall, these results highlight the important role of interactions in enabling the QME within our four-site Bose--Hubbard model. While the non-interacting system exhibits conventional relaxation governed by initial proximity to equilibrium, on-site interactions introduce nonlinear coupling between coherence, entanglement, and dissipation. This coupling reshapes the relaxation spectrum and drives anomalous thermalization dynamics. Although our study is restricted to a small system, the observed behavior suggests that many-body correlations may play a crucial role in the emergence of QME in larger dissipative bosonic systems, motivating further investigation beyond the four-site regime.

\subsection{Clean Bose--Hubbard Regime with External Potentials}
\label{subsec:bhm_external}

Having established the critical role of on-site interactions in facilitating the QME in Sec.~\ref{subsec:clean_bhm}, we now investigate whether spatial inhomogeneity, introduced via external potentials, can serve as an alternative mechanism for anomalous relaxation dynamics, following the procedure outlined in Sec.~\ref{sec:method}. Unlike interactions, which directly modify many-body correlations, external potentials such as a linear Stark potential or random on-site disorder primarily affect the single-particle energy landscape while preserving the quantum many-body structure. This distinction, contrasted with the non-interacting and interacting clean regimes in Secs.~\ref{subsec:clean_tb} and \ref{subsec:clean_bhm}, prompts the question: can spatial symmetry breaking induce the nonlinear relaxation pathways necessary for QME?

We first consider a linear Stark potential, which introduces controllable site-to-site energy differences, creating a competition between gradient-induced localization and kinetic and thermal delocalization. We implement the procedure using a four-site bosonic chain with an equal number of particles, maintaining the on-site interaction and low hopping strength established in Sec.~\ref{subsec:clean_bhm} to operate in the Mott insulating regime. Initial thermal states are prepared at a fixed inverse temperature, each characterized by a distinct Stark potential strength, and evolved under Lindblad dynamics with local dephasing noise acting on particle number operators. All systems converge toward a common reference steady state defined by a specific Stark potential strength. Relaxation dynamics are monitored using four quantum metrics: trace distance $D[\hat{\rho}(t), \hat{\rho}_{ss}]$ for geometric distinguishability, relative entropy $S[\hat{\rho}(t) || \hat{\rho}_{ss}]$ for information-theoretic divergence, entanglement asymmetry $\Delta S_A$ for a chosen bipartition to assess particle number symmetry restoration, and $\ell_1$-norm of quantum coherence $\mathcal{C}[\hat{\rho}(t)]$ in the Fock basis to capture quantum superposition dynamics.

\begin{figure*}[t]
\centering
\includegraphics[width=0.80\linewidth]{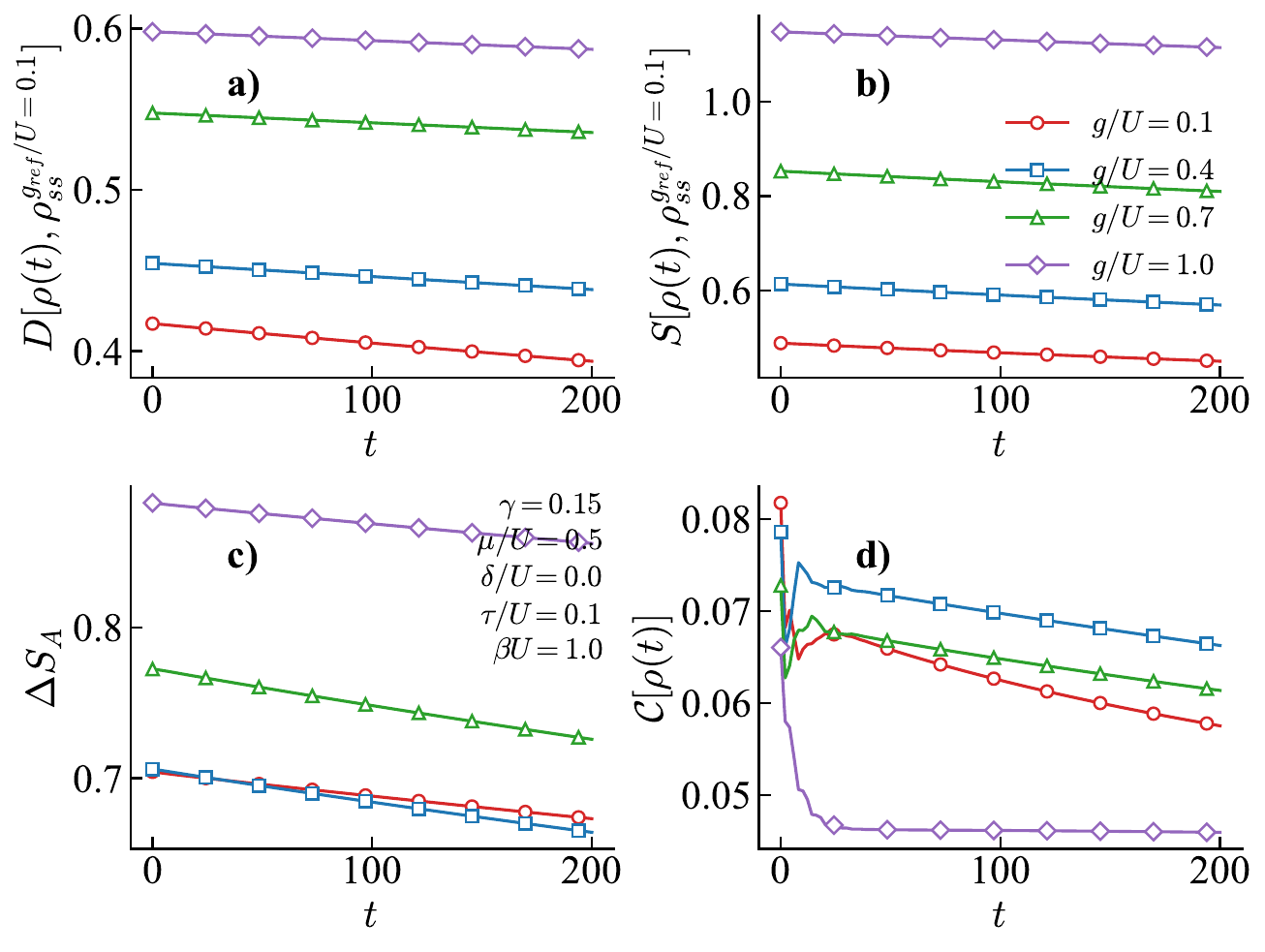}
\caption{Suppression of relaxation dynamics under Stark potential. Time evolution of quantum metrics for a Bose--Hubbard system with varying Stark field strengths $g$. A four-site system with fixed on-site interaction evolves from thermal initial states under Lindblad dynamics with local dephasing noise (rate $\gamma/U$). The reference steady state $\hat{\rho}_{ss}$ corresponds to a fixed Stark strength $g_\text{ref}$. All four metrics show systematic retardation of thermalization: \textbf{(a)} trace distance $D[\hat{\rho}(t), \hat{\rho}_{ss}]$ exhibits maintained hierarchy with increased separation, \textbf{(b)} relative entropy $S[\hat{\rho}(t) || \hat{\rho}_{ss}]$ shows delayed convergence for stronger potentials, \textbf{(c)} entanglement asymmetry $\Delta S_A$ for bipartition $A$ demonstrates persistent symmetry breaking, and \textbf{(d)} $\ell_1$-norm of quantum coherence $\mathcal{C}[\hat{\rho}(t)]$ reveals coherence suppression under spatial inhomogeneity. Increasing Stark potential strength creates localization barriers that impede thermalization and prevent QME emergence. 
}
\label{fig:gradient_dynamics_low_tau}
\end{figure*}

The results, shown in Fig.~\ref{fig:gradient_dynamics_low_tau}, contrast sharply with the QME observed in the clean interacting regime. The trace distance in panel~\ref{fig:gradient_dynamics_low_tau}(a) exhibits a hierarchical ordering, with stronger Stark potentials maintaining greater distance from the steady state throughout the evolution, indicating systematic thermalization impediments absent in Sec.~\ref{subsec:clean_bhm}. The relative entropy in panel~\ref{fig:gradient_dynamics_low_tau}(b) corroborates this monotonic convergence hierarchy, showing slower information-theoretic equilibration for stronger potentials, unlike the crossover behavior previously observed. Entanglement asymmetry in panel~\ref{fig:gradient_dynamics_low_tau}(c) reveals persistent symmetry breaking under stronger potentials, contrasting with the efficient symmetry restoration enabled by interactions alone. The quantum coherence dynamics in panel~\ref{fig:gradient_dynamics_low_tau}(d) further indicate suppression of quantum superpositions, suggesting rapid decoherence or quantum coherence retention due to the inhomogeneous potential landscape, distinct from the interaction-driven quantum coherence pathways in Sec.~\ref{subsec:clean_bhm}. These findings suggest that Stark-induced spatial inhomogeneity inhibits QME by introducing localization barriers, unlike the anomalous relaxation facilitated by interactions.

\subsection{Disordered Bose--Hubbard Regime without Stark Potential}
\label{subsec:disordered_bhm}
To further explore the impact of spatial inhomogeneity, we investigate random on-site disorder in the absence of a Stark potential, maintaining the on-site interaction and low hopping strength from the previous cases. Initial thermal states are prepared at a fixed inverse temperature, each with a distinct disorder amplitude, and evolved under Lindblad dynamics toward a common reference steady state defined by a specific disorder strength. The same four quantum metrics are used to track relaxation dynamics.

Following the exploration of Stark potentials in Sec.~\ref{subsec:bhm_external} and building on the interaction-driven QME observed in Sec.~\ref{subsec:clean_bhm} and the non-interacting baseline in Sec.~\ref{subsec:clean_tb}, we investigate whether random on-site disorder can induce the QME in the Bose--Hubbard model, adhering to the procedure outlined in Sec.~\ref{sec:result}. By maintaining zero Stark potential ($g/U = 0$), fixed on-site interaction, and low hopping strength in the Mott insulating regime, we isolate the effects of random disorder-induced translational symmetry breaking. This setup tests the universality of QME across different physical mechanisms, exploring whether disorder can overcome thermalization bottlenecks, akin to the interaction-driven pathways in Sec.~\ref{subsec:clean_bhm}, or if it mirrors the localization effects seen with Stark potentials.

\begin{figure*}[t]
\centering
\includegraphics[width=0.80\linewidth]{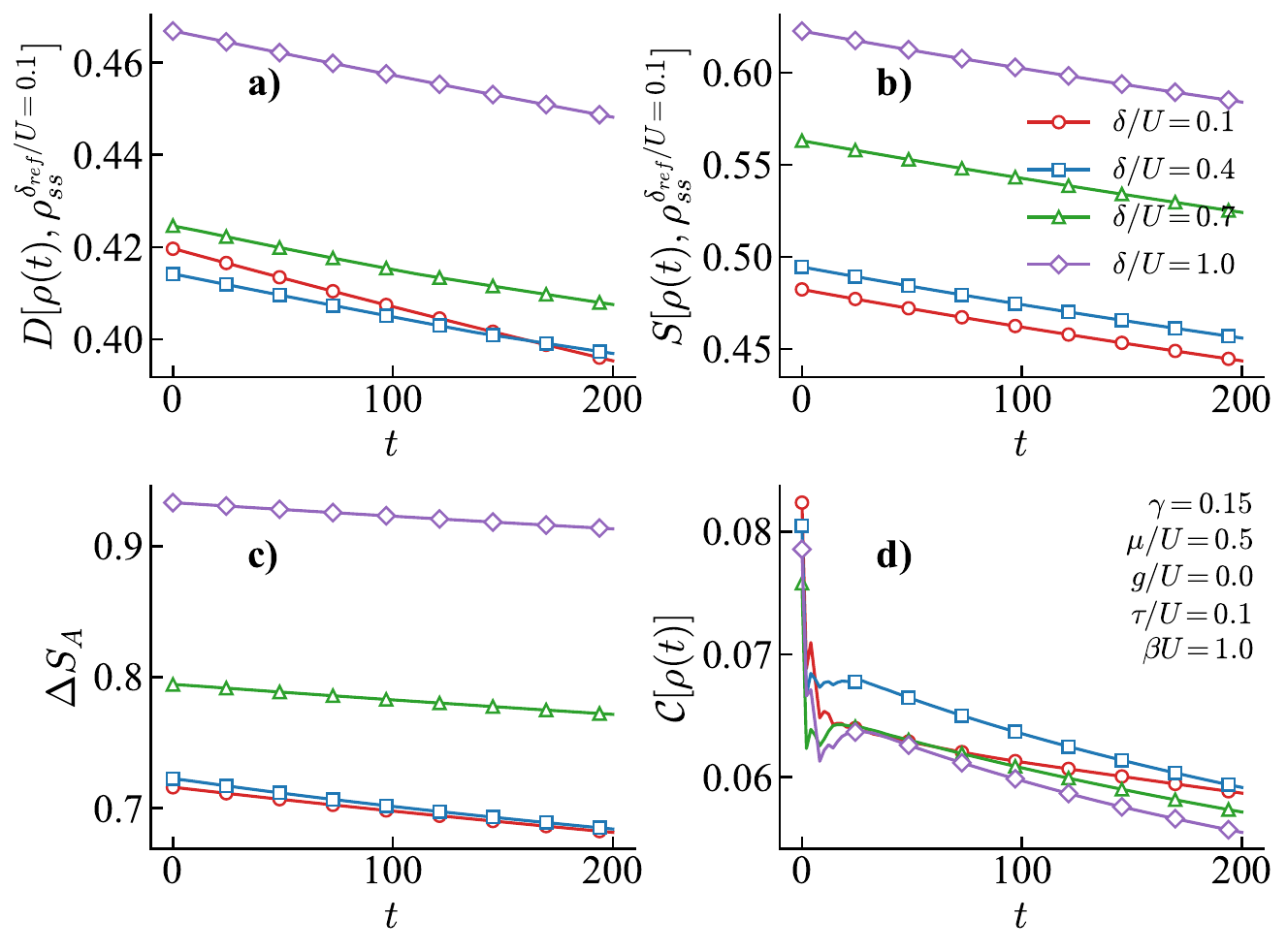}
\caption{Moderate thermalization suppression under random disorder. Relaxation dynamics of a Bose--Hubbard system with varying disorder amplitudes $\delta$. A four-site system with fixed on-site interaction evolves from thermal initial states under Lindblad dynamics with local dephasing noise (rate $\gamma$). The reference steady state $\hat{\rho}_{ss}$ corresponds to a fixed disorder amplitude $\delta_\text{ref}$. All metrics indicate mild localization effects: \textbf{(a)} trace distance $D[\hat{\rho}(t), \hat{\rho}_{ss}]$ shows modest convergence delays, \textbf{(b)} relative entropy $S[\hat{\rho}(t) || \hat{\rho}_{ss}]$ exhibits maintained relaxation hierarchy, \textbf{(c)} entanglement asymmetry $\Delta S_A$ for bipartition $A$ reflects disorder-induced symmetry breaking, and \textbf{(d)} $\ell_1$-norm of quantum coherence $\mathcal{C}[\hat{\rho}(t)]$ demonstrates slower coherence decay. While random disorder impedes thermalization compared to the clean interacting case, its effects are less severe than Stark potential-induced localization, and no QME signatures are observed. 
}
\label{fig:superlattice_dynamics}
\end{figure*}

Figure~\ref{fig:superlattice_dynamics} illustrates that random disorder suppresses QME signatures, akin to the Stark potential case in Sec.~\ref{subsec:bhm_external}, but with milder effects compared to the pronounced localization observed there. The trace distance in panel~\ref{fig:superlattice_dynamics}(a) shows a modest but consistent convergence delay with increasing disorder amplitude, maintaining a hierarchical ordering, unlike the crossover events seen in the clean interacting regime (Sec.~\ref{subsec:clean_bhm}). The relative entropy in panel~\ref{fig:superlattice_dynamics}(b) corroborates this, indicating reduced relaxation efficiency due to mild localization effects, less severe than the Stark potential’s impact but similar to the conventional thermalization in Sec.~\ref{subsec:clean_tb}. Entanglement asymmetry in panel~\ref{fig:superlattice_dynamics}(c) reveals disorder-induced translational symmetry breaking, with higher disorder amplitudes sustaining greater asymmetry, though less persistently than with Stark potentials. The quantum coherence dynamics in panel~\ref{fig:superlattice_dynamics}(d) exhibit slower decay at intermediate to late times with increasing disorder, consistent with inhibited thermal mixing, yet lacking the dramatic suppression seen with strong Stark potentials.  
The short-time features in $\mathcal{C}[\rho(t)]$ (Figs.~\ref{fig:gradient_dynamics_low_tau}(d) and \ref{fig:superlattice_dynamics}(d)) arise from the separation of timescales between: (i) rapid suppression of off-diagonal Fock-basis elements by the dephasing dissipator (rate $\gamma$), and (ii) slower population redistribution induced by the Hamiltonian. This produces an initial dephasing kink (a fast drop or transient) before the dynamics enters a slower transport-limited regime, especially pronounced when Stark gradients or disorder inhibit hopping-assisted mixing.

Our Stark and disorder observations align qualitatively with localization physics in tilted/disordered Bose--Hubbard chains, where gradients or randomness suppress transport and thermalization \cite{kolovsky2007bloch,kolovsky2008dynamical,Fischer2016,hooley2019stark}. In those studies, tilt-induced Wannier--Stark localization can be stronger and more systematic than moderate disorder, consistent with our finding that increasing $g$ produces a more dramatic retardation than comparable $\delta$ in the explored parameter range. From the Liouvillian perspective, inhibited transport reduces effective mixing between number configurations, enhancing slow-mode dominance and suppressing the possibility of relaxation-order inversion.
These findings indicate that random on-site disorder disrupts thermalization by introducing mild localization effects, but it does not induce the nonlinear relaxation pathways necessary for QME, unlike the interaction-driven crossovers in Sec.~\ref{subsec:clean_bhm}. Compared to the Stark potential’s stringent localization barriers, disorder imposes less severe constraints, yet both external potentials contrast sharply with the anomalous dynamics enabled by many-body correlations. This emphasizes the necessity of interaction-driven mechanisms for QME emergence in dissipative bosonic systems.

\section{Conclusion}
\label{sec:conclusion}

We investigated relaxation-order inversion (quantum Mpemba effect) in an open Bose--Hubbard chain described by a Markovian Lindblad master equation with local number dephasing. Using exact evolution in the fixed-$N$ sector of a four-site, unit-filled lattice, we compared families of thermal initial states evolved under a common reference generator and quantified their approach to the stationary state using trace distance, quantum relative entropy, a symmetry-projected entropy imbalance (often termed entanglement asymmetry), and $\ell_{1}$-norm of quantum coherence.

In the clean interacting regime, all metrics exhibit crossings: initial states that are farther from the stationary state at $t=0$ can converge faster at late times. Within the Liouvillian-mode framework, this behavior is naturally interpreted as an interaction-induced redistribution of initial overlaps with the slow decay subspace, enabling reduced weight on the slowest-decaying mode(s) for some initially distant thermal preparations. In contrast, the non-interacting limit shows monotone relaxation ordering across all diagnostics, consistent with the absence of interaction-driven reshaping of slow-mode weights.

External inhomogeneities suppress QME in the explored parameter windows. A linear Stark field produces the strongest retardation of relaxation, consistent with Wannier--Stark localization and transport suppression, which enhances slow-mode dominance and prevents ordering inversion. Random on-site disorder yields milder delays than the Stark field but similarly maintains a fixed relaxation hierarchy without crossings. Short-time coherence transients are explained by rapid dephasing of off-diagonal elements followed by slower, transport-limited population mixing.

Our results highlight that, in dephasing-driven open bosonic chains, on-site interactions are the key ingredient enabling relaxation-order inversion, while strong localization mechanisms suppress it. Extending these studies to larger lattices and directly tracking Liouvillian gaps and slow-mode weights versus system size and inhomogeneity strength are promising next steps toward a finite-size scaling theory of QME in interacting lattice bosons.
\\
\\
\textbf{Data availability:}
 All data generated or analyzed during this study are included in this published article.\\
\\
\textbf{Competing interests:}
 The authors declare no competing interests.\\
 \\
\textbf{Acknowledgments:}
This work was supported by the Qatar National Research Fund (QNRF) under grant number ARG01-0603-230468.  H.A.Z. and M. J. acknowledge funding by the Slovak Research and Development Agency under the contract No. APVV-24-0091, and by the grant of The Ministry of Education, Research, Development and Youth of the Slovak Republic under the contract No. VEGA 1/0298/25.
H.A.Z. acknowledges the financial support provided under the postdoctoral fellowship program of P. J. \v{S}af\'{a}rik University in Ko\v{s}ice, Slovakia.\\

\bibliography{bibliography} 

\end{document}